%% file: ms.tex
\documentclass{aa}
\usepackage{txfonts}
\usepackage{graphicx}
\usepackage{natbib}
\usepackage{xspace}
\bibpunct{(}{)}{;}{a}{}{,}

\input{macros}

\input{definitions.tex}

\usepackage{color}

\begin{document}

% The following information is for internal review, please remove them for submission
%\widetext
%\leftline{Version as of \today}
%\centerline{\em INTERNAL DOCUMENT -- NOT FOR PUBLIC DISTRIBUTION}

% the following line is for submission, including submission to the arXiv!!
%\hspace{5.2in} \mbox{Fermilab-Pub-04/xxx-E}

%\title{Probing dark cosmology and gravity with the cell count moments of bright galaxies}

\title{Determination of the abundance of cosmic matter \\
via the cell count moments of the galaxy distribution}

\author{J. Bel\inst{1,2} \& C. Marinoni\inst{1,2,3}}
\institute{Aix Marseille Universit\'e, CNRS, Centre de Physique  Th\'eorique, UMR 7332, F-13288, Marseille, France
\and Universit\'e de Toulon, CNRS, CPT, UMR 7332, F-83957 La Garde, France
\and Institut Universitaire de France, 103, bd. Saint-Michel, F-75005 Paris, France}

\authorrunning{Bel \& Marinoni}

\date{Received --; accepted --}

\abstract{
We demonstrate that accurate and precise information about the matter content of the universe can be retrieved 
via a simple cell count analysis  of the 3D spatial distribution of
galaxies. A new clustering statistic, the  {\it galaxy clustering
  ratio} $\eta$, is the key to this process.
This is defined as the ratio between  one- and two-point second-order moments of the smoothed galaxy density distribution.
The distinguishing feature of this statistic is its universality: on large cosmic scales both galaxies (in redshift space) and mass (in real space)
display the same $\eta$ amplitude. This quantity, in addition,  does not evolve as a function of redshift.
As a consequence, the $\eta$ statistic provides insight into characteristic  parameters of the real-space 
power spectrum of mass density fluctuations  without the need to
specify the galaxy biasing function,  neither a model for galaxy redshift distortions, nor the  growing mode of  density ripples. 
We demonstrate the method with the luminous red galaxy (LRG) sample extracted from the spectroscopic Sloan Digital Sky Survey (SDSS) data release 7 (DR7) catalogue. 
Taking weak (flat) priors  of the curvature  of the   universe ($\Omega_k$) and of the  constant value of the dark energy 
equation of state ($w$), and strong (Gaussian) priors  of the physical baryon density $\Omega_bh^2$, of the Hubble constant $H_0$, and of the spectral index of primordial density perturbations $n_s$, 
we estimate the abundance of matter  with a relative error of $8\%$ ($\Omega_m= 0.283\pm0.023$).
We expect that this  approach  will be instrumental in  searching for evidence of new physics beyond the standard model of cosmology and in 
planning future redshift surveys,  such as  BigBOSS or EUCLID.
}


\maketitle

\section{Introduction}

Determining the value of the constitutive parameters of the Friedman equations
is a problem of great observational and  interpretative difficulty, 
but owing to its bearing on fundamental physics,  
it is one of great theoretical  significance.  Current estimations suggest that 
we live in a homogeneous and isotropic  universe where baryonic matter is a minority (1/6) of all matter, 
matter is a minority (1/4) of all forms of energy, geometry is spatially flat, and
cosmic expansion is presently accelerated  \citep[e.g.][]{mod1, mod3,mod4,
san, anderson, mariso, papPlanck}.

However evocative these cosmological results may be, incorporating  them
into a physical theory of the  universe is a very perplexing problem.
Virtually all the attempts to  explain the nature of dark matter which
is the  indispensable ingredient 
in models of cosmic structure formation \citep{fw},  as well as  of
dark energy which is the physical mechanism that drives 
cosmic acceleration \citep{peeb},  invoke exotic physics beyond current theories \citep{feng, cope, clif}. 
Measurements, on the other hand,  are not   
yet precise enough to exclude most of these proposals \citep{amen12}. The Wilkinson microwave anisotropy probe (WMAP, \citet{mod1}) and the Planck mission \citet{papPlanck}, 
for example,   fix the parameters of a `power-law $\Lambda$CDM' model with impressive accuracy. 
This is a cosmological model characterized by a flat geometry, by a positive  
dark energy (DE) component $\Omega_X$ with $w=-1$,  and by primordial perturbations that are scalar,  
Gaussian, and adiabatic \citep{cosmo}.
A combination of astrophysical probes are needed
if we are to constrain deviations from this minimal model.  
Because they are sensitive to different sets of 
nuisance parameters and to different subsets of  the full cosmological
parameter set, different techniques provide
consistency checks, lift parameter degeneracies, and enable stronger 
constraints and, in the end, a safer theoretical interpretation \citep{frie}.

A way to meet this challenge is by designing and exploring the potential
of new cosmic  probes.  In this spirit, we have recently proposed
to exploit  the internal dynamics of  
isolated disc galaxies  \citep{mardiscs} and of pairs of galaxies \citep{marpairs} in order to set limits on relevant cosmological parameters.
Here we demonstrate that precise and accurate cosmological information 
can be extracted from the large scale spatial distribution of galaxies if 
their clustering properties  are characterized through  the measurement of the clustering ratio $\eta$,  the ratio between the
correlation function, and the variance of the galaxy overdensity field smoothed on a scale $R$.

Gaining insight into the dark matter and dark energy sectors
via the analysis of the 3D clustering of galaxies is a tricky task.
There is no reason to expect (and there is observational evidence to the contrary) that the galaxies
trace the underlying matter distribution exactly. A fundamental
problem is thus understanding how to map the clustering of
different galaxy types into the clustering of the underlying matter,
which the theories most straightforwardly predict.

Most of the attempts to address this issue centre on the feasibility
of reconstructing the biasing relation from independent
observational evidence, for example weak lensing surveys (thus
increasing the complexity of the observational programes) or on
the possibility of considering biasing as a nuisance quantity that
can be statistically marginalized (thus increasing the number of
parameters of the model and degrading the predictability of the
theory). The  complementary line of attack taken  in this paper
consists in developing  right from the start a bias-free statistical descriptor of clustering, 
that is an observable that can be directly compared to theoretical predictions. 

The paper is organized as follows. We  define the clustering ratio $\eta$ in \S 2,  and we discuss its theoretical  properties  in \S 3.  In \S 4 we test its robustness and performances 
by analysing  numerical simulations of the large scale structure of the universe. Cosmological constraints obtained from SDSS DR7 sample \citep{sdss}  are derived and discussed 
in \S 5. Conclusions are drawn  in section 6.

Throughout, the Hubble constant is parameterized via $h =H_0/100$ km
s$^{-1}$Mpc$^{-1}$. Data analysis
is not framed in any  given  fiducial cosmology;  i.e., we compute the clustering ratio statistic in any tested
cosmology.

\section{A new cosmological observable:\\ the galaxy clustering ratio}

We characterize the inhomogeneous distribution of galaxies 
in terms of the  local dimensionless density contrast  
\begin{equation}
\delta_g({\bf x}) \equiv \frac{ \rho_g({\bf x})}{\langle\rho_g({\bf x})\rangle}-1,  
\end{equation}
where $\rho_g({\bf x})$ is the comoving density of galaxies in real space,
and where $\langle\rho_g({\bf x})\rangle$ denotes the selection
function, which is the expected 
mean number of galaxies at position ${\bf x}$, given the selection criteria of the survey.
Since the galaxy distribution is a stochastic point process, a 
spherical top-hat filter $W$ of radius $R$ is applied to generate a continuous,  
coarse-grained observable 
\begin{equation} 
\delta_{g,R}({\bf x})=\int \delta_g({\bf y}) W(|{\bf x-y}|/R)d{\bf y}.
\end{equation}
The second-order, one-point  
\begin{equation} 
\kappa_{20,g,R}=\langle \delta_{g,R}^2({\bf x}) \rangle_c
\end{equation}
and  the two-point   
\begin{equation} 
\kappa_{11,g,R}({\bf r}) = \langle \delta_{g,R}({\bf x})\delta_{g,R}({\bf x} +{\bf r}) 
\rangle_c
\end{equation}
cumulant moments are the lowest order,  non-zero connected moments of the 
probability density functional (PDF)  of the field $\delta_{g,R}({\bf x})$ 
\citep{szasza1, szasza2, szasza3, berna, berna02, bm}. If the PDF 
is stationary  and isotropic, the ratio $\kappa_{11,g,R}({\bf r})/\kappa_{20,g,R}$  
is equivalent to the one between the correlation 
function and the  variance of the filtered field  
\begin{equation} 
\eta_{g,R}(r,{\bf p})=\frac{\xi_{g,R}(r,{\bf p})}{\sigma^2_{g,R}({\bf p})},
\label{etagr}
\end{equation}
\noindent where we have explicitly emphasized  the dependence of the observable 
on the set ({\bf p}) of cosmological parameters. This comes through because the statistical descriptor (\ref{etagr}) 
can be estimated from  data only  once a  comoving distance-redshift
conversion model has been supplied.

\begin{figure}
\includegraphics[scale=0.55]{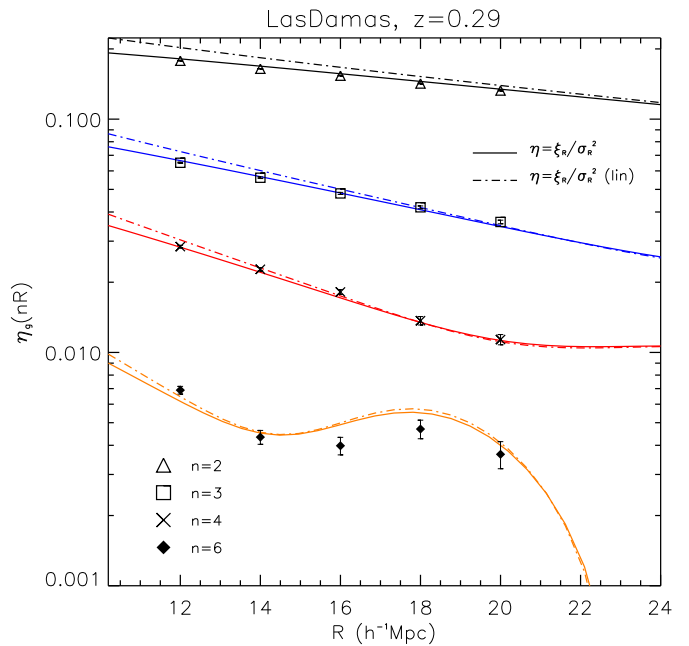}
\caption{\label{fig:sim} {\it Left panel:}
The redshift-space galaxy clustering ratio $\eta_{g,R}(nR)$ (LHS of eq. \ref{core}) extracted from  160 independent LasDamas simulations  of the distribution of luminous red galaxies (LRG)  (see \S 4). 
The observable $\eta_{g,R}(nR)$ is shown as a function of the smoothing radius $R$ for  four different values of $n$ as indicated in the inset.
Simulated observations  are compared to the predicted  amplitudes of the real-space mass clustering ratio $\eta_R(nR)$ (RHS of eq. (\ref{core}))
for the same values of $n$.   Dot-dashed lines show predictions obtained using  a linear power spectrum,  while solid lines are derived 
using   the non-linear (halofit) model of  \citet{smith}.   Both  the galaxy and mass clustering ratios  are derived by adopting  the same cosmological parameters of the LasDamas simulations.
Error bars represent the uncertainties expected from a redshift survey covering a volume 160 times larger than that explored by the LRG sample used in this study.}
\end{figure}

\subsection{Estimating the galaxy clustering ratio}

The variance  $\sigma^2_{g,R}$ and the correlation function $\xi_{g,R}$
of a smoothed density field of galaxies  can be estimated  from the  three-dimensional distribution of galaxies  following the procedure outlined in \citet{bm}.
We assume that the random variable $N$  models the number of galaxies  within  typical
spherical cells (of constant comoving radius $R$) and consider the  dimensionless galaxy excess  
\begin{equation}
\delta_N\equiv\frac{N}{\bar N}-1,
\end{equation}
where $\bar N$ is the mean number of galaxies contained in the cells.

To estimate the one-point second-order moment of  the galaxy overdensity field ($\kappa_{2} = \langle \delta_N^2 \rangle_c$),  we fill the survey volume with the maximum number ($n_t$) of 
non-overlapping spheres of radius $R$ (whose centre is called  a {\it seed}) and we compute 
\begin{equation}
\hat\kappa_{2}=\frac{1}{n_t}\sum_{i=1}^{n_t}\delta_{N,i}^2,
\end{equation}
\noindent where $\delta_{N,i}$ is the dimensionless counts excess  in the $i$-th sphere. 

The  two-point second-order moment $\kappa_{11}=\langle \delta_{Ni}\delta_{Nj}\rangle_c$ follows from 
a generalization of this cell-counting process. To this purpose,  
we add a motif of isotropically distributed spheres  around each seed and retain as {\it proper seeds}  only  
those for which the spheres of the motifs lie completely   within the survey boundaries. 
The centre of each new sphere is separated from the proper seed  by the length $r=nR$ (where $n$ is a
generic real parameter usually taken to be an integer without loss of
generality), and the pattern is designed in such a way as to maximize the number of quasi non-overlapping spheres 
at the given distance $r$. In this study, as in Bel and Marinoni (2012),  the maximum  allowed  overlap  between contiguous spheres  is set to $2\%$ in volume. 
This value represents a good comprise between the need to maximize the number of the spheres that are correlated to a given seed 
and the volume probed by them. If there is a substantial overlap of the spheres of the motif, all the signal on the correlation scales r is fully sampled. 
However, this strategy is  computationally expensive. On the other hand, if the number of spheres is too small,  then a substantial fraction of the information available at a given 
correlation scale $r$ is lost.  We have verified that varying this  threshold in the range $2$\%- $50$\% does not modify the estimation of one- and two-point statistical properties of the counts.

An estimator of the average excess counts in the $i$ and $j$ cells  at separation $r$ is 
\begin{figure}
\includegraphics[scale=0.55]{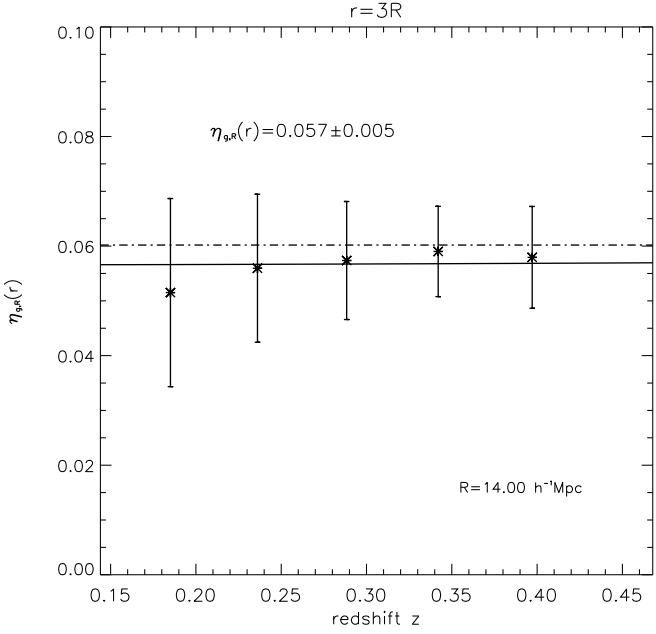}
\caption{\label{fig:simred} 
The redshift-space galaxy clustering ratio $\eta_{g,R}(nR)$  extracted from the LasDamas simulations  (see \S 4)  using $R=14h^{-1}$Mpc and $n=3$
is shown as a function of  redshift. Markers represent the average
measurement from  40  LasDamas  mock catalogues, each of them
simulating a redshift survey that has the 
same  radial  density profile  and observational selection effects as the LRG sample of SDSS DR7.
Simulated observations  are compared to the predicted  amplitudes of the real-space mass clustering ratio $\eta_R(nR)$ (right-hand side of eq. (\ref{core}).   
Dashed lines show predictions computed  using a linear power spectrum,   while solid lines are derived 
assuming    the non-linear (halofit) model of  \citet{smith}.   Both
the galaxy and mass clustering ratios   are estimated by adopting the same cosmological parameters as the LasDamas simulations.
Error bars represent the  uncertainty from measurements  in a single mock catalogue.  They are obtained as the standard deviation of 40
independent measurements,  therefore including the contribution due to  cosmic variance.}  
  \end{figure}
\begin{equation}
\hat\kappa_{11}=\frac{1}{n_t n_{mot}}\left\lbrace\sum_{i=1}^{n_t}\delta_{N,i}\sum_{j=1}^{n_{mot}}\delta_{N,ij}\right\rbrace,
\end{equation}
\noindent where  $n_{mot}$ is the number of spheres in the motif,   and where $\delta_{N,ji}$ is the excess count in the  
motif's cell $j$ at distance $r$ from the seed $i$.
  
It is natural, or at least convenient,  to model the spatial point distribution of galaxies  as a process resulting from the discrete sampling of an underlying continuous stochastic 
field $\lambda_g({\bf x})$. The quantity of effective physical  
interest to which we want to have access is thus
$\delta_{g,R}({\bf  x})\equiv \Lambda_g({\bf x}) / {\bar \Lambda_g}-1$, where $\Lambda_g({\bf x}) =\int_{V({\bf x})} \lambda({\bf x'})d{\bf x'}$ is the 
continuous limit of the discrete counts  $N$ in volume V. As a consequence,   it is  necessary to correct counting estimators for discreteness effects.
To this purpose and following standard practice in the field, we model the sampling as a  local Poisson process  (LPP,  \citet{Layser}). Specifically,  
we  map moments of the discrete variable N into moments of its continuous limit by using 
\begin{equation}
\langle\Lambda_g^k\rangle=\langle N(N-1)...(N-k+1)\rangle \equiv \langle(N)_f^k\rangle,
\label{unpoint}
\end{equation}
\noindent in the case of  one-point statistics,  and its generalization \citep{szasza2, bm}
\begin{equation}
\langle\Lambda_g^k({\bf  x}_1)\Lambda_g^q({\bf  x}_2)\rangle =  \langle(N_1)_f^k(N_2)_f^q\rangle,
\label{deuxpoint}
\end{equation}
\noindent for the  two-point  case. As a result, the  estimators of  the variance and the correlation function, corrected for shot noise effects, 
 are \citep{szasza1, bm}
\begin{eqnarray}
\hat \sigma_{g,R}^2 &= &\hat\kappa_{2}-\bar N^{-1}   \nonumber  \\
\hat \xi_{g,R}  &=   &\hat\kappa_{11} .
\label{ccum}
\end{eqnarray}

By construction,  the $\eta$  observable is  not sensitive to the shape of the radial  selection function, if the density gradient 
is nearly constant on scales $r \sim 2R$. Sample-dependent corrections are  mandatory, instead,  if the survey geometry is not regular; i.e., a significant fraction of the  
counting cells  overlap the survey boundaries. Bel et al.  (2013), for example, show how to  minimize the impact of radial 
(redshift) and angular incompleteness  when the $\eta$  estimator  is applied to a survey, the VIMOS Public Extragalactic Redshift Survey \citep[VIPERS,][]{guz13},  characterized by 
non-trivial selection functions.

The scaling of the galaxy clustering ratio  $\eta_{g,R}(nR)$ as a function of both $R$ and $n$ is shown in figure \ref{fig:sim}. 
That galaxies tend to cluster on small cosmic scales is revealed by the fact that  
the amplitude of $\eta$  decreases monotonically as a function of the  correlation scale  $r$.
In other terms,  the  probability of finding two cells  with density contrasts  of equal sign is suppressed 
as  the separation between the cells increases.   
The trend of $\eta_{g,R}(r)$ as a function of $R$ (for a given fixed value of $r$) is the opposite, implying
 that the relative loss of power  that results from  filtering the field  on larger and larger scales   is stronger   
in  one-point than in two-point statistics (at least when second-order moments are considered.)  
We also note that,  for constant values of $n$, $\eta_{g,R}(nR)$ decreases monotonically as a function  of the smoothing radius $R$ 
only if  the field is correlated on small scales, i.e. $n<4$.  
For higher $n$ values,  in fact,  the clustering ratio becomes sensitive to the  baryon acoustic oscillations (BAO) 
imprinted on the large scale distribution of galaxies. As a consequence,  a peak shows up at  $n= l/R$, where $l$ is the position of the BAO relative maximum in the correlation function
of galaxies.

\section{Theoretical predictions for the amplitude of the clustering ratio}
 
It is straightforward to predict the theoretical  value of the 
second-order galaxy clustering ratio $\eta_{g,R}$. 
On large enough cosmic scales $R$, matter fluctuations  are small
and are described by the linear  power spectrum
\begin{equation}
\Delta^2(k, {\bf p}) = 4\pi A k^{n_s+3}T^2(k, {\bf p}),
\label{plin}
\end{equation}
where   $A$ is a normalization factor, $n_s$ the primordial spectral index,  
and $T^2$ the transfer function. (In this study we assume the analytic approximation 
of  \citealt{EH}.) Accordingly, 
the amplitudes of the second-order statistics for mass evolve as a function 
of redshift ($z$) and scale ($R$) as 
\begin{equation} 
\sigma^2_R(z,{\bf p})=\sigma^2_8(z=0)D^2(z){\mathcal F}_R({\bf p}), 
\end{equation}
and
\begin{equation} 
\xi_R(r,z, {\bf p})=\sigma_8^2(z=0) D^2 (z){\mathcal G}_R(r, {\bf p}).
\end{equation}
 
Both these equations are normalized on the scale  $r_8=8h^{-1}$Mpc,  
where $D(z)$ represents  the linear growing mode \citep{cosmo},  
while the effects of filtering are incorporated in the functions
\begin{equation}
{\mathcal F}_R({\bf p})=\frac{\int_0^{+\infty}\Delta^2(k, {\bf p})\hat{W}^2(kR)d\ln k}{\int_0^{+\infty}\Delta^2(k,{\bf p})\hat{W}^2(kr_8)d\ln k}
\label{ff}
\end{equation}
\begin{equation}
{\mathcal G}_R(r, {\bf p})=\frac{\int_0^{+\infty}\Delta^2(k, {\bf p}) \hat{W}^2(kR) j_{0}(kr) d\ln k}{\int_0^{+\infty}\Delta^2(k, {\bf p})\hat{W}^2(kr_8)d\ln k}
\label{gg}
\end{equation}
\noindent where $j_{n}(x)$ is the spherical Bessel function of order $n$, and $\hat{W}$ is the Fourier transform of the window function.  In this analysis we adopt a spherical  top hat  filter for which
\be
\hat{W}(kR)=\frac{3}{kR}j_1(kR).
\ee

By analogy with the galaxy clustering ratio, we can now define the mass clustering ratio 
as 
\begin{equation}
\eta_{R}(r, {\bf p})=\frac{{\mathcal G}_R(r, {\bf p})}{{\mathcal F}_R({\bf p})}.
\label{etar}
\end{equation}

The relationship between the mass and galaxy clustering ratios in real space
follows immediately once we specify how well the overall matter distribution 
is traced by  its luminous subcomponent on a given scale $R$. On large
enough scales, 
such as those explored in this paper,  a  local, deterministic,  non-linear biasing scheme \citep{fg}, namely 
\be 
\delta_{g,R}({\bf x})=\sum_{i=0}^{N}(b_{i,R}/i!)\delta_R^i({\bf x}) , 
\label{nlbias}
\ee
offers a fair description of the mapping between mass and galaxy density fields. 
The scale-dependent parameters $b_{i,R}$ are called biasing parameters,  and within the  
limits $\sigma_{R}^2<<|b_{1,R}/b_{2,R}|$  and $\xi_{R}<<1/b_2$  (see   eq. 28 of Bel \& Marinoni 2012), 
one has
\ba
\sigma^2_{g,R}(\vr) & \approx  &  b_{1,R}^2 \sigma^2_{R}(\vr)  \nonumber \\
\xi_{g,R}(\vr)  & \approx &    b_{1,R}^2 \xi_{R}(\vr).   
\label{biassim}
\ea
We therefore deduce that 
\begin{equation}
\eta_{g,R}(r, {\bf p}) = \eta_R(r, {\bf p}), 
\label{core}
\end{equation}
\noindent a relation independent of the specific value of the bias amplitudes $b_{1R}$.

The conditions under which eq. (\ref{core}) is derived are quite generic and are 
fulfilled once the field is filtered on sufficiently large scales $R$ 
and/or  once the second-order bias coefficient is negligible compared to the linear bias term. This last
requirement is satisfied for  large  $R$  (for example,  Marinoni et al. (2005, 2008b), using VVDS galaxies at $z=1$  \citep{lefevre}, 
find $|b_{2,R}/b_{1,R}|  \sim  0.17 \pm0.07$ for $R=10 h^{-1}$Mpc, a result implying  that, at least on these scales,  any eventual scale dependence of the biasing 
relation is also negligible.

\begin{figure*}
\begin{center}
\includegraphics[scale=0.55]{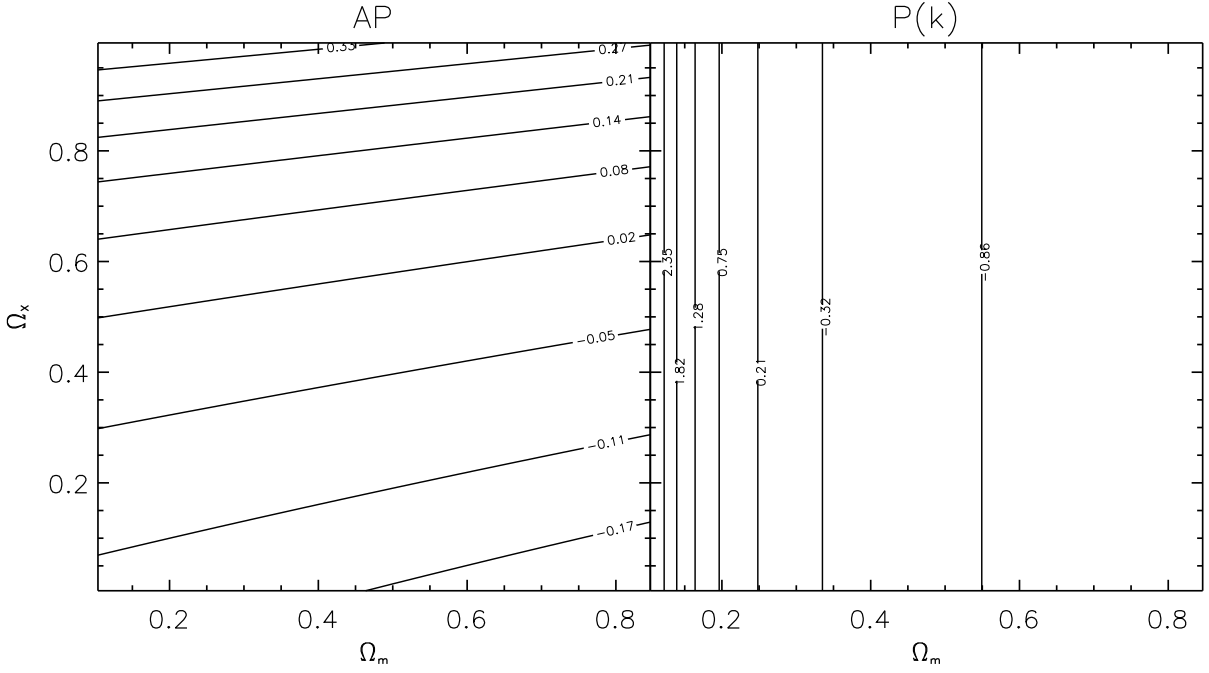}
\caption{\label{fig:appk} 
{\it Left:} Isocontours of the function $\eta_{g,R}/\eta_{g,R}^{best fit}-1$ displaying the 
relative variation in the galaxy clustering ratio (with respect to the best-fitting value deduced from the analysis of SDSS DR7 data) 
induced by a  wrong choice of the  distance-redshift conversion model  (i.e. a wrong guess of the cosmological parameters $[\Omega_m, \Omega_X]$). 
The galaxy clustering  ratio is reconstructed on  the scales $(R,r)=(14,42)h^{-1}$Mpc, and its expression is approximated well by eq. (\ref{appeta}).
The best-fitting cosmology  inferred from the analysis of SDSS DR7 data is $\Omega_m=0.28, \Omega_X=0.45, w=-1.2$.
The isocontours  measure the strength  of the AP geometric distortions.  {\it Right:}   the relative variation 
in the  mass clustering ratio   $\eta_{R}/\eta_{R}^{best fit}-1$  that results from  computing the power spectrum in the wrong cosmology.}
\end{center}
\end{figure*}

\noindent 

The most interesting aspect of  eq. (\ref{etar}) is that, in  the linear limit,  it is effectively insensitive
to redshift distortions, and, therefore, independent of their specific modelling. In other terms,   
 the ratio $\eta_{R}(r, {\bf p})$ between second-order statistics has identical 
amplitude  in both real- and redshift-space. If the only net effect of peculiar 
velocities is to enhance the amplitudes of the density ripples in Fourier space, without any change in 
their phases or frequencies (as predicted, for example, by the Kaiser
model, \citealt{kaiser}),  then redshift distortions contribute  equally  to  
the numerator and denominator of eq. (\ref{etar}).  This simplifies the interpretation of the master equation (\ref{core}): its
left-hand side (LHS) can be estimated using galaxies in redshift space,  while its right-hand side (RHS) can be theoretically evaluated in real space. 
This factorization property holds because the $\eta$ statistic is only sensitive to the monopol of the power spectrum,  i.e. to the quantity  
obtained by averaging the  redshift-space anisotropic power spectrum  $P_z({\bf k})$ over angles in k space ($\int_{-1}^{1} P^z(k,\mu_k)d\mu_k$,
where $\mu_k$ is the cosine of the angle between the $k$ vector and the line-of-sight). Indeed, as explained in section 2.1, 
to estimate $\eta$,  we position an isotropic distribution of cells (the motif) around each seed.

Another salient property  is that,  as long as a linear matter's  power spectrum is assumed, 
the mass clustering ratio $\eta_{R}$ is  effectively independent of the amplitude of the  
matter power spectrum normalization parameter $A$.  In linear regime,  i.e. when $R$ is sufficiently large, the clustering ratio is
also independent of redshift. This property is graphically illustrated in Fig. \ref{fig:simred}. This comes about because   the only time dependence appearing in the expression of the 
variance and correlation function of a smooth density field is through the growing mode $D(z)$, a quantity that, on large linear scales, appears in the relevant 
equations as a multiplicative parameter that eventually factors out in the $\eta$ ratio.  Figure \ref{fig:simred} shows that,  on large scales and sufficiently low redshift, this property still holds even when 
a non-linear  description of the matter power spectrum is adopted. Specifically,  it shows that the  mass clustering ratio  predicted using the 
non-linear power spectrum  of  \citet{smith}  is effectively time independent. In conclusion, for any given filtering ($R$) and  correlation ($r$) scales,
the mass clustering ratio $\eta_{R}(r, {\bf p})$ behaves as a cosmic `standard of clustering', a  quantity that does not evolve across cosmic time. 

The bias free property of  eq. (\ref{core}) also deserves further comment.
Since the smoothing and correlation  are performed on  fixed scales $R$ and $r$, $\eta_{g,R}$ is not a function but a number. 
Although one can estimate it using  any given visible tracer of the large scale structure of the universe, this number captures second-order information about 
the clustering of the general distribution of mass. This conclusions holds as long as eq. \ref{nlbias}  provides a fair description of the  nature of biasing 
on large scales $R$, which is as long as the matter - galaxy relation  is deterministic and local. 
 In this regard we remark that the galaxy
density is likely to show some scatter around the dark matter density
due to various physical effects. This stochasticity, however,  is not expected to  influence the 
accuracy of  the equation (\ref{core}) in a significant way. On large scales,  the only effect stochasticity might introduce   is a rescaling of the linear biasing parameter by a constant  \citep{scherrer}. 
Because it is defined as  a  ratio,  $\eta$  is unaffected by such a systematic effect.  
A second source of concern comes from the fact  that  the $\eta$ formalism is built out of the hypothesis that biasing is local; i.e., we can neglect
the possible existence of   scale dependent operators in the matter-galaxy density relations, at least on large smoothing scales $R$.
A simple argument provides the guess for the sensitivity of the clustering ratio to this assumption.
 In the limit in which $r>R$,  $\hat{W}^2(kR)j_0(kr)\sim j_0(kr)$, the mass clustering ratio scales approximatively
 as
 \be
 \eta_R(nR,{\bf p}) \propto \frac{\Delta^2(k_{j_0}, \bf{p})}{\Delta^2(k_{W_0},{\bf p})}, 
 \label{simp}
 \ee
where, quite crudely,  $k_{j_0}$ and $k_{W_0}$ can be estimated as the
wave vectors  where  the amplitude of the low pass filters  $j_0(kr) $
and $\hat{W}^2(kR)$  falls to one half.
The proposed test, therefore, exploits neither  the absolute amplitude nor  the full shape of the power spectrum, but only its relative strength on two distinct
$k$ scales. By choosing to correlate the field on scales $r$ that are not too large compared to the smoothing scale $R$, it is possible to minimize the impact of any eventual  $k-$dependent bias.

\begin{figure*}
\begin{center}
\includegraphics[scale=0.55]{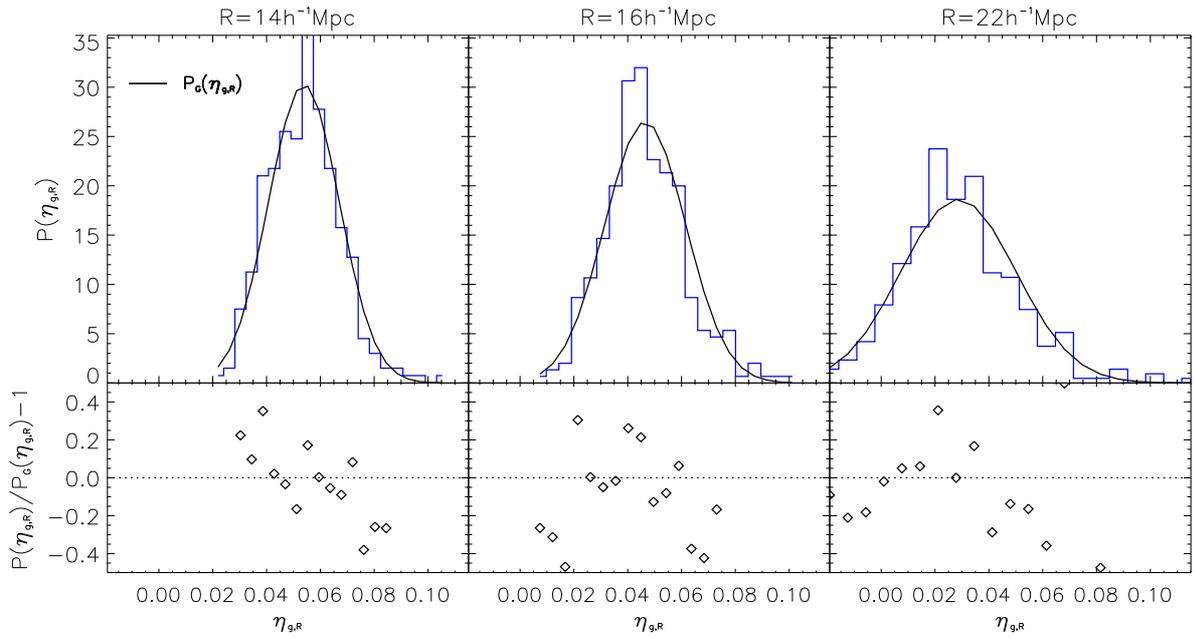}
\caption{\label{fig:histo} 
{\it Upper panels:} the distribution  of the galaxy clustering ratio $\eta_{g,R}(r)$ obtained by analysing 160 LasDamas  mock catalogues  simulating the distribution of  luminous red galaxies
 (histograms). The observable $\eta_{g,R}(r)$ is computed  for the  three different smoothing radii $R$ shown on top of each plot and by assuming, furthermore, that  $r = 3R$. 
The solid line displays the best-fitting Gaussian PDF. {\it Lower panels:}  the relative deviation between the Gaussian model and the actual distribution of the data is shown.}
\end{center}
\end{figure*}

\subsection{Precision and accuracy of the $\eta$ formalism}

The next step is to make sure  that  systematic effects,   whether physical (such as any  eventual non-locality in the biasing relation), observational 
(like systematic errors in  estimating  the relevant observables  from data),  or statistical (that,  for example, although the estimator in eq. (\ref{etagr}) is a ratio of unbiased estimators, 
it does not necessarily to be unbiased itself) do not compromise the effectiveness of  the formalism in practical applications.
We therefore use numerical simulations of the  large scale structure to compare  the amplitude of the  redshift space galaxy clustering  ratio (the LHS of  eq. (\ref{core})) 
and  the theoretically predicted  value of the real space mass  clustering ratio 
(the RHS of eq. (\ref{core})).  Figure \ref{fig:sim}  shows that, on scales $R>10h^{-1}$Mpc and $r=3R$, equation (\ref{core}) is good to better than 
$1 \%$,  and to better than $0.1\%$ on scales larger than 15$h^{-1}$Mpc (for the same correlation length $r=3R$).   This last figure is two  orders of magnitude 
smaller  than the precision achievable when measuring $\eta_{g,R}(r)$ using current data ($\sim 7\%$ see Sec IV).
Simulations also indicate that,  for smoothing scales  $R<15h^{-1}$Mpc,  the precision achieved by using a non-linear power spectrum (in our case the phenomenological 
prescription of  \citealt{smith}) is nearly a factor of five  better
than the one  obtained by  adopting a simple  linear model. 

It is impressive how this remarkable sub-percentage precision in   mapping theory onto real-world observations is achieved 
without introducing  any external non-cosmological  information, such
as any explicit biasing parameter or  any model  that corrects galaxy positions for 
non-cosmological redshift distortions.  Nonetheless, for very large separations (i.e. $r=6R$), theoretical predictions 
fail to reproduce observations. As a matter of fact, when the field is smoothed on a scale $R=16h^{-1}$Mpc and correlated on a scale  $r=16 \times 6 \sim 100h^{-1}$Mpc, 
the $\eta$ indicator becomes extremely sensitive to the specific features  of the baryon acoustic oscillations   imprinted in the galaxy distributions.
Indeed, from a theoretical point of view, the correlation function of the smoothed density field results from convolving the correlation function of 
galaxies over two smoothing windows of radius $R$ 
at  positions ${\bf x}_1$ and ${\bf x}_2$ separated by the distance ${\bf r}$
\begin{equation}
\xi_R({\bf r})=\frac{1}{V_R^2}\int_{V_{R({\bf x}_1)}}d{\bf y}_1\int_{V_{R({\bf x}_2)}}d{\bf y}_1\xi({\bf y}_1,{\bf y}_2).
\end{equation}
This convolution preserves the position of the BAO peak, but it  decreases its amplitude and  broadens its width.
Notwithstanding, it is still possible  to detect  the BAO scale in a precise way using the eta estimator.  As the poor agreement between theory and measurements 
 of Fig. \ref{fig:sim} shows, however,    the  simple power spectrum models described above are not able to grasp the non-linear physics involved in this phenomenon.  
The modelling of  local peculiar motions is also paramount if we are to predict  the BAO profile to the precision required for  cosmological purposes.

At the opposite limit,  when the field is correlated  on small scales $(r=2R)$, the mass clustering ratio fits observations in a better way if  the non-linear prescription of \citet{smith} is considered instead  of the linear power spectrum.  
In other terms, numerical simulations indicate that once the  $\delta-$field is coarse grained on a sufficiently large scale  $R$, 
a linear power spectrum captures the essential physics  governing the clustering of galaxies  only  if the field is correlated on scales  $3 \leq n<6$.

\subsection {The clustering ratio as a cosmic probe}

Apart from the most immediate use as a statistic  to measure the clustering of mass in a universal (sample-independent) way, 
the clustering ratio can also serve as a tool to set limits to the value of cosmological parameters.
The RHS of eq. (\ref{core}) relies upon the theory of  cosmological perturbations,  and it is fully specified (it is essentially analytic) given the shape of the mass power spectrum in real space. 
Therefore it is directly sensitive to 
the shape ($n_s$) of the primordial power spectrum of density fluctuations, as well as to  the parameters
determining the shape of the transfer function of matter perturbations at matter-radiation equality
(in particular, the present-day extrapolated value of the   matter  ($\Omega_m h^2$) and  baryon  
($\Omega_b h^2$) density parameters. The possible contribution of massive 
neutrinos is neglected in this analysis.)  
By contrast,   the LHS term of eq. (\ref{core}) probes the structure of the  comoving distance-redshift
relation, therefore it is fully  specified once the  homogeneous expansion rate history of the universe is known. On top of $\Omega_m$, the LHS term is thus also sensitive to  $\Omega_X$ and  $w$, 
{\it i.e.} to a wide range of DE models. If distances are expressed in units 
of $h=H_0/100$ km s$^{-1}$Mpc$^{-1}$, the LHS is effectively independent of  the value of the Hubble 
constant $H_0$. 

The possibility of constraining the true cosmological model  follows from noting that the equivalence expressed by eq. (\ref{core}) holds true if and only if the LHS and RHS are both
estimated in  the correct cosmology. Analytically,  the amplitudes of the galaxy clustering ratio $\eta_{g,R}(r)$ and of the  mass clustering ratio  $\eta_R(r)$ do not match  if  clustering
is analysed by assuming the  wrong cosmology; since the derivatives  with respect to the parameters ${\bf p}$ of the LHS and RHS  of eq. \ref{core} are not  identical. 
Physically, two effects contribute to breaking the equivalence  expressed by eq. \ref{core} when a wrong set of parameters ${\bf p}$ is assumed: 
geometrical distortions, the Alcock and Paczynski (AP) signal \citep{ap, bph, marpairs}, 
dynamical effects, and distortions in the power spectrum shape. By  AP,   we mean the artificial anisotropy  in the clustering signal that results when the density 
field is correlated using the wrong distance-redshift relation.  In this case, and contrary to what one would expect on the grounds of symmetry considerations,  
the clustering power in the  directions parallel and perpendicular to the observer's line of sight is not  statistically identical. 
Transfer function shape distortions, the second source of sensitivity to cosmology, are generated by evolving  the relevant  mass statistics in the wrong cosmological background,  
and they can be  traced primarily to the spurious modification of the size of the horizon  scale at the epoch of equivalence.

  The  amplitude of AP and power spectrum distortions are compared in Fig. \ref{fig:appk}.  
  Geometric effects (for  $R=14 h^{-1}$Mpc and  $r=3R$),  
  are almost completely degenerate with respect to the $\Omega_m$ parameter, that is  for any given fixed value of 
  $\Omega_X$, the amplitude of the AP signal  is nearly  independent of $\Omega_m$. On the contrary, the distortions in the observable $\eta_{g,R}$ induced by a wrong guess of the dark energy parameter  
  are an increasing function of $\Omega_X$: the amplitude of the clustering ratio is underestimated(/overestimated) by at most $20\%$ when $\Omega_X$ varies in the interval $[0, 1]$. 
  Power spectrum distortions, instead, which  are by definition  independent of  $\Omega_X$,  are extremely sensitive to the matter density parameter $\Omega_m$. A change of $\sim \pm 0.1$ of 
  the true value (in this case $\Omega_m=0.28$) results in a change of $\eta_R$ by nearly  $\pm 50\%$.
  We conclude that,   given the uncertainty  characterizing the SDSS data
  analysed in this paper, the AP signal is  marginal -- but it will become an interesting
  constraint on dark energy with larger datasets, such as Euclid \citep{euc}.

\begin{figure*}
\begin{center}
\includegraphics[scale=0.5]{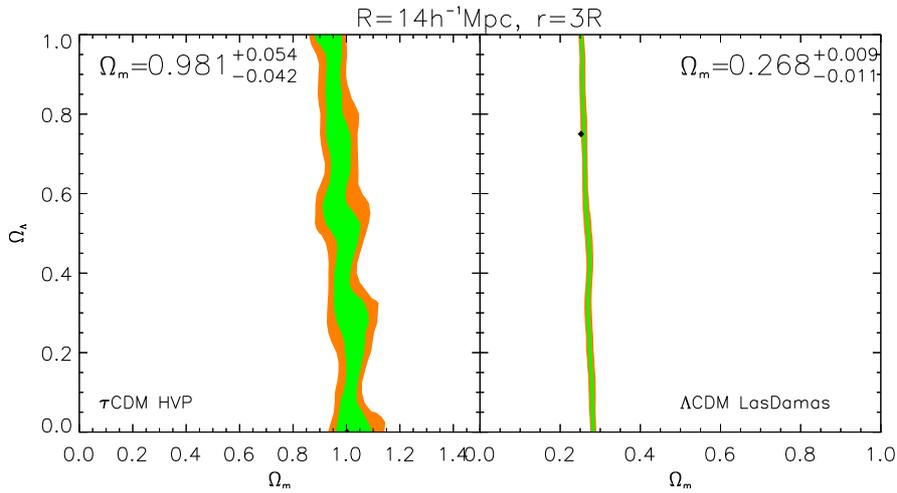}
\caption{\label{fig:liksim} {\it Left panel:}
two-dimensional confidence limits on $\Omega_m$ and $\Omega_{\Lambda}$
from  a `blind' analysis of the $\tau$CDM HVP simulation \citep{HVP}. 
Contours are plotted for ${\mathcal L}/{\mathcal L}_{min}< 2.3, 6.17$ corresponding 
to $68$  and $95$ per cent {\it c.l.} for a multivariate Gaussian 
distribution with 2 degrees of freedom.  
The relevant one-dimensional marginalized constraint is shown in the legend.
The clustering ratio $\eta_{g,R}(r,{\bf p})$ is estimated using 
$R=14h^{-1}{\mathrm Mpc}$ and $r=3R$. For the best-fitting cosmology we measure
$\eta_{g,R}(r, {\bf p}^{f})=0.0294\pm0.0017$ where the  
error is evaluated via $30$ jackknife resampling of the data, 
excluding each time a sky area of $36\times 19.5$ deg$^2$.
Dirac delta  priors are taken of $\Omega_b h^2$, $H_0$, and $n_s$, that   
are centred on the simulated values.
{\it Right panel:} same as before, but  now 
contours represent the joint likelihood analysis of $40$ independent $\Lambda$CDM 
mock catalogues simulating  the SDSS LRG sample. 
For the best-fitting cosmology we find $\eta_{g,R}(r, {\bf p}^{f})=0.0573\pm 0.0008$, where the error, computed as the 
{\it s.d.} of the mean of $40$ measurements,  includes the contribution from cosmic variance.}
\end{center}
\end{figure*}

\begin{figure}
\begin{center}
\includegraphics[scale=0.55]{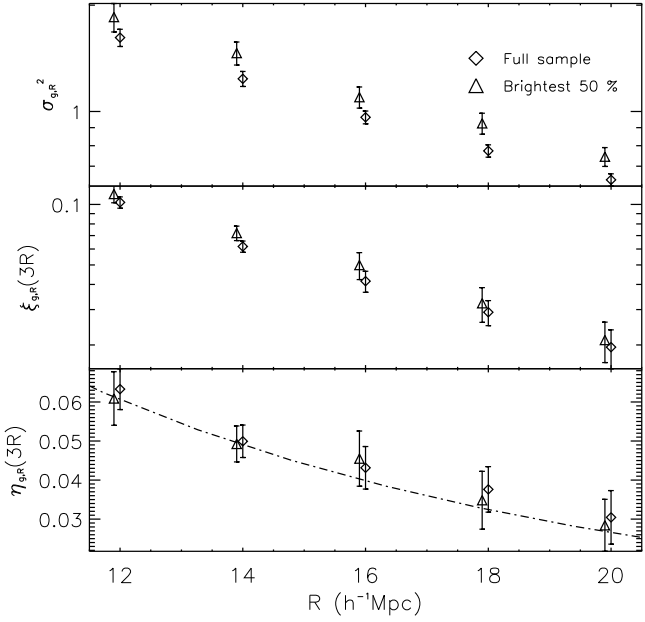}
\caption{\label{fig:eta} 
{\it Upper:}  the variance  $\sigma_{g,R}=\xi_{g,R}(0)$  of the SDSS LRG sample as a function of the filtering scale $R$
for 2 different sets: the full sample (diamonds)  and the  subsample containing the brightest $50\%$ (in the $r$ band). 
{\it Centre:} the correlation function  
$\xi_{g,R}(3R)$ of the smoothed galaxy density field  is shown as a function of $R$ for the same subsets.
Both the variance and the correlation function  are estimated in the best-fitting cosmological model.   
  {\it Lower: }  the amplitude of the galaxy clustering ratio $\eta_{g,R}(3R)$  is demonstrated  to be the same for these two samples of biased tracers of the 
 large scale distribution of matter. Also shown is the theoretically  predicted scaling of the mass clustering ratio  (dot-dashed line). This last  quantity 
 is computed assuming  the  best-fitting 
cosmological parameters ${\bf p}^f =(0.28, 0.45, -1.2, 73.8, 0.0213, 0.96)$ deduced  from analysis of the whole LRG sample  on the specific scale $R=14h^{-1}$Mpc. 
In all the panels, error bars  are derived  from 30 block-jackknife resampling of the data, 
excluding, each time, a sky area of $12\times 14$ deg$^2$.}
\end{center}
\end{figure}

\begin{figure*}
\begin{center}
\includegraphics[scale=0.75]{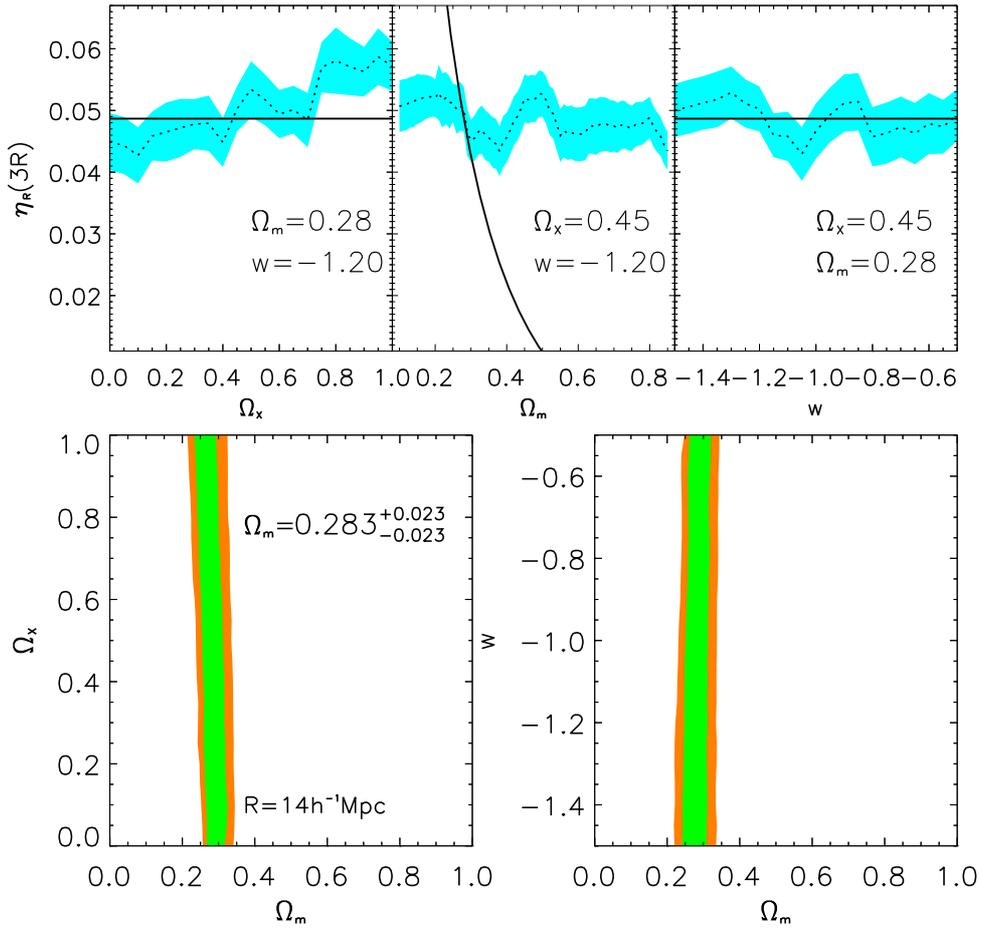}
\caption{\label{fig:data} 
{\it Upper panels:} The galaxy clustering ratio $\eta_{g,R}(r,\Omega_m, \Omega_X, w)$ 
(dotted line)  reconstructed from SDSS DR7 data assuming $R=14h^{-1}$Mpc and $r=3R$, 
 plotted as a function of $\Omega_{\Lambda}$ (left), $\Omega_m$ (centre) and $w$ 
(right), after fixing the remaining cosmological  parameters, shown in the inset, 
to the values that minimize the logarithmic posterior (\ref{logpost}). 
Grey shading is the $1-$sigma deviation around the mean, and it is  
computed from 30 block-jackknife resampling of the data.   This uncertainty depends on cosmology (because of the redshift-distance conversion model)
in a stochastic way ($\pm 20\%$ around the mean, for the cosmologies shown in the picture).                                                                             
We also show  the predicted scaling of the mass clustering ratio 
(solid line) in the corresponding cosmology.
The best set of cosmological parameters  ${\bf p}^f= (0.28, 0.45, -1.2, 73.8, 0.0213, 0.96)$  is the very one  that minimizes the 
difference between theoretical predictions (line)  and actual data (points). For SDSS DR7 data,
we find $\eta_{g,R}(3R, {\bf p}^{f})=0.0502\pm 0.0036$ on a scale $R=14h^{-1}$Mpc. 
{\it Lower panels:} two-dimensional marginalized constraints 
on a curved XCDM model in which both $\Omega_X$ and $w$ are allowed to vary.
Gaussian priors are taken of $\Omega_b h^2=0.0213\pm 0.0010$, of  
$H_0=73.8\pm 2.4$ km s$^{-1}$Mpc$^{-1}$ and of $n_s=0.96\pm 0.014$ from BBN \citep{bbn}, HST \citep{hst} and 
WMAP7 \citep{lars} determinations, respectively.
Contours are plotted for ${\mathcal L}/{\mathcal L}_{min}< 2.3,6.17$.
}
\end{center}
\end{figure*}

\begin{figure}
\includegraphics[scale=0.55]{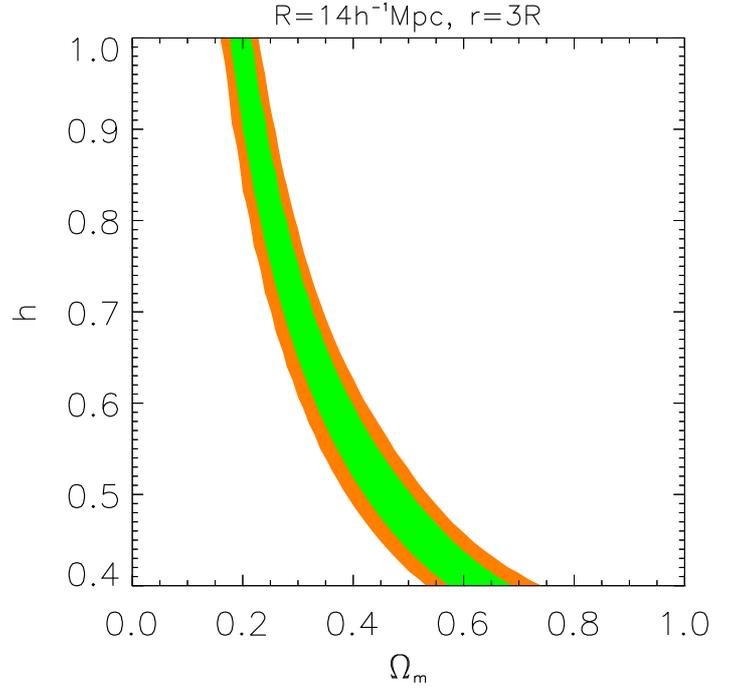}
\caption{\label{fig:degeneracy} 
Two-dimensional marginalized constraints on $\Omega_m$ and $H_0$ obtained by fitting data with a curved XCDM model.
Contours are plotted for ${\mathcal L}/{\mathcal L}_{min}< 2.3,6.17$
Gaussian priors are taken of $\Omega_b h^2=0.0213\pm 0.0010$ and $n_s=0.96\pm 0.014 $ from BBN \citep{bbn} and WMAP7 \citep{lars} determinations, respectively.}
\end{figure}

We  now illustrate how, in practice, we  evaluate  $P({\bf p}|\eta_{g,R})$,  the  likelihood of the unknown set of parameters
 ${\bf p}=(\Omega_m, \Omega_{X}, w, H_0, \Omega_bh^2, n_s)$  given the actual value of the observable 
$\eta_{g,R}$. The  analysis does not require any specification of further 
model parameters other than those quoted above.

The probability distribution function of the clustering ratio is not  immediately obvious since this observable is 
 defined via a ratio of two non-independent random variables. Simulated and real data  suggest  that it is fairly accurate  
 to assume that  the PDF of $\eta_{g,R}$ is approximately Gaussian. This  is  demonstrated in Fig. \ref{fig:histo}. 
 As a result, the  most likely set  of cosmological parameters (${\bf p}^{f}$) is the one that minimizes the logarithmic posterior 
${\mathcal L}=-\log P$ 
\begin{equation}
{\mathcal L}= \sum^{N} \log \sigma_{\eta,i} +\frac{\chi^2}{2}- \log \pi +B, 
\label{logpost}
\end{equation}
\noindent where $\chi^2=\sum^{N}\sigma_{\eta,i}^{-2}(\eta_{g,R}-\eta_R)^2$, where $N$ is the 
number of estimates of $\eta_{g,R}$, $\sigma_{\eta,i}$ is the error on the observable, 
$\pi$ describes any a-priori information about the PDF of ${\bf p}$,  
and where $B$ is a normalization constant that can be fixed by requiring 
$\int Pd{\bf p}=1$. 

Since  eq. (\ref{core}) 
is free of look-back time effects, the analysis does not require slicing the sample in
arbitrary redshift bins. Only one estimation ($N=1$) of eq. (\ref{etagr})  
is needed across the whole sample volume. 
We have recalculated the observable $\eta_{g,R}(r, {\bf p})$
for each comoving distance-redshift model, {\it i.e.} on a grid 
$(\Omega_m, \Omega_X, w)$ of spacing $[10^{-2}, 5 \cdot 10^{-2}, 5 \cdot 10^{-2}]$. 
As a consequence, the posterior ${\mathcal L}$
does not vary smoothly between different  models because the number of 
galaxies counted in any given  cell varies from model 
to model. However, since computing the  observable $\eta_{g,R}(3R, {\bf p})$  
takes a limited amount of time, shot noise is the price we have decided to pay 
to obtain an unbiased likelihood hyper-surface.

\section{Blind analysis of cosmological simulations}

We assess the performances of the $\eta$  test  under realistic operating conditions 
via a `blind' analysis of mock catalogues that are characterized 
by widely different sets of expansion rate parameters,  power spectrum shapes,
mass tracers, and  radial selection functions. These are the $\tau$CDM Hubble Volume 
Project (HVP) simulation \citep{HVP} and the $\Lambda$CDM  LasDamas simulation \citep{lasdamas}. 

The HVP is a synthetic catalogue of clusters of galaxies  that covers one octant of sky, 
extends over the redshift interval $0.1<z<0.43$,  is comprised of $\sim 10^6$ massive haloes 
with an average space density of $6\times 10^{-4}h^3{\mathrm Mpc}^{-3}$, and 
it was simulated using ${\bf p}^{i}=(1,\; 0,\; -1,\; 21,\; 0,\; 1)$. 
%The radial density profile of this halo sample is shown in Fig. \ref{fig:den}. 
An end-to-end analysis of clusters mock data (HVP) allows us to check for any insidious 
algorithmic biases that could arise when training a method to `recognize' only 
a fiducial $\Lambda$CDM cosmology via a single class of biased tracer 
of the matter clustering pattern, namely galaxies. The left-hand panel of  Fig. \ref{fig:liksim} shows that 
the input value $\Omega_M=1$ is statistically retrieved.

The  $\Lambda$CDM  LasDamas simulations are a set of 40 nearly independent  galaxy simulations  that 
cover $120\times 45$ deg$^2$,   extend over the redshift interval $0.16<z<0.43$, are  comprised of $\sim 60000$ 
galaxies  each,  and have an average space density of $\sim 8.9\times 10^{-5}h^3{\mathrm Mpc}^{-3}$. 
The mock catalogues are obtained from N-body simulations (whose input cosmological parameters are ${\bf p}^{i}=(0.25,\; 0.75,\; -1,\; 70,\; 0.0196,\; 1)$)
by populating dark matter haloes with galaxies according to a halo
occupation distribution  function. The structural parameters of this statistic,  which describes the probability distribution of the
number of galaxies in a halo as a function of the host halo mass, are fixed using  SDSS data. 
These catalogues,  which incorporate all the observing selections of the SDSS luminous red galaxies 
 survey, i.e. the real data sample used in our analysis (see \S 5),  allow us to verify that the specific SDSS observing biases  do not spoil our cosmological inferences. 
 Another advantage is that  we can  forecast the statistical (and systematic) fluctuations in the observable and compare this figure, which includes contribution from cosmic variance, 
 with  the uncertainty in $\eta$ directly estimated  from SDSS DR7 data via a block-jackknife technique (see \S 5). 
 
The encouraging outcome of the analysis of these SDSS like data  is presented in Fig. \ref{fig:liksim}, where we show
the cosmological bounds that could be obtained from a  redshift survey  having  a volume 40 times larger than is probed by the 
SDSS DR7 LRG sample  ($\sim 1/2$ of the volume that  will be measured by 
surveys such as EUCLID \citep{euc} or BigBOSS \citep{boss}). This figure shows that one 
can constrain $\Omega_m$ precisely 
($\sim 3.7\%$) and accurately (the input value is recovered within the
$95\%$ c.l.).  Whereas a proper analysis of an EUCLID-like simulation
is beyond the purposes of the present work, our   clustering analysis   in a comparable  volume  is already suggestive 
of the robustness  with which future data will allow constraining the matter density parameter. As Fig. \ref{fig:simred}  indicates,   
pushing the technique to its limit, at least on a scale $R=14h^{-1}Mpc$, only requires incorporating predictions  from
a  non-linear matter power spectrum. 
Finally, the strong sensitivity to the abundance of matter  of the clustering ratio  essentially arises  because
the zero-order spherical Bessel function in eq. (3) 
filters different portions of $\Delta^2$  when $\Omega_m$ is changed.
If we bias high  $\Omega_m$, the  suppression of power in $\xi_R(r,{\bf p})$ on a scale $r$ 
is more than what is observed when estimating $\xi_{g,R}(r, {\bf p})$ using the 
corresponding wrong distance-redshift relation (see Fig. \ref{fig:appk}).

\section{Cosmological constraints from the SDSS DR7 sample}

We now present and discuss cosmological constraints inferred from the  analysis of  the LRG sample extracted from the SDSS DR7.

The geometry of the subsample that we have analysed is dictated by the need of sampling the galaxy  distribution  with cells 
of radius $R$, as well as of correlating cell counts on scales $r$.  The inferior limit on $R$ is set as to simplify our analysis.  
Although not mandatory,  by framing the analysis in the linear domain, i.e. by choosing an inferior threshold  for $R$,  we
avoid introducing any phenomenological description of the matter power spectrum. 
It is true that  this choice prevents us from extracting the maximum information from the data;  however, it 
makes the  data analysis more transparent. 
For example,  by adopting  as a model  the quasi linear model (halofit) of \citet{smith} for the non-linear power spectrum, 
 we  would be forced to add an extra component, $\sigma_8(0),$  to the vector of  unknown  parameters ${\bf p}$, 
therefore complicating unnecessarily the likelihood analysis and the marginalization procedures. 
The choice of $R$ is additionally conditioned by the practical requirement of   minimizing the shot noise contribution in 
each cell, {\it i.e.} $R>(4\pi \rho/3)^{-1/3}$.  In this work we adopt the  scale $R=14h^{-1}$Mpc. 
The lower limit $r=3R$ on the correlation scale, instead, is set  to guarantee   
optimal accuracy in the approximation (\ref{core}).
At the opposite end, the higher the values of 
$R$ and $r$, the less the inferred cosmological predictions are informative.
The  sample that complies with these constraints covers the redshift interval 
$0.15<z<0.43$, has a contiguous sky area of $120\times 45$ deg$^2$, 
and is comprised of $62,652$ LRG (corresponding to 55\% of the
  total number of LRG contained in DR7),  
with a mean density of $9.2\times 10^{-5}h^3{\mathrm Mpc}^{-3}$.

The  $R-$scaling  of  the  variance  $\sigma_{g,R}$ and  correlation function  $\xi_{g,R}(nR)$ of the smoothed galaxy density field is shown 
in Fig.  \ref{fig:eta}.   The amplitude of both  these observables  is not a universal quantity. In addition to the background cosmological model,  
it also depends on the  particular set of luminous objects used to trace the mass density field.  That  bright galaxies display 
stronger $rms$ fluctuations and are more strongly correlated than faint ones as shown in Fig. \ref{fig:eta}. 
This same figure also shows that the  amplitude of $\xi_{g,R}$ is nearly one order of magnitude smaller than $(\sigma_{g,R})$.
The lack of power in the two-point statistic  is theoretically  expected, since only in the limit for $r \rightarrow 0$ does the correlation function 
of a smoothed field converge to the variance, i.e. $\xi_{g,R}(0)=\sigma^2_{g,R}$. Given that in a sample of $N$ cells,  the  number  of independent combinations of  two cells 
is smaller than $N$,  we also expect the two-point statistic to be  estimated with  larger uncertainties
than the corresponding one-point statistic of equal order.  For example, the SDSS DR7 sample  allows us to extract the value of  $\xi_{g,R} (/\sigma^2_R)$  
with a  precision of $6.5\%(/5\%)$ when the field is  smoothed on the scale $R=14h^{-1}$Mpc. 
As far as the  functional dependence on $R$ is concerned,  the scaling
of both  these  statistics is essentially featureless and approximated  well by a power law over the interval $12<R<20 h^{-1}$Mpc.
For  a given fixed value of $n$,  the relative loss of power resulting from filtering the field on larger and larger scales $R$ is stronger  for $\xi_{g,R}(nR)$  than  $\sigma^2_{g,R}$.  

 The galaxy clustering ratio $\eta_{g,R}(nR)$ estimated in the best-fitting cosmology is shown in Fig. \ref{fig:eta}.  This observable 
 is a universal quantity associated to the mass density field, and,  as such, it is independent of the specific biasing properties of the mass tracers adopted in this analysis.
The  precision  with which the clustering ratio is determined from SDSS LRG data  ($\sigma_{\eta}/\eta  \sim 7\%$  
on a scale $R=14h^{-1}$Mpc)   is  estimated  from 30 jackknife
resampling of the data each time  excluding  a sky area of $12 \times 14$ deg$^2$. This figure
is in excellent  agreement with the one   ($\sim 8\%$) deduced from the  analysis of the standard deviation  displayed  by the $40$ SDSS-like simulations LasDamas,
which include, by definition,  the contribution from cosmic
variance. This indicates that $\eta$, which is defined as a ratio of
equal order statistics and thus contains the same stochastic source,
is weakly sensitive to this systematic effect. We note that the relative uncertainty  on  $\eta_{g,R}(nR)$ cannot be simply deduced via standard propagation of errors since measurements of $\sigma^2_{g,R}$ and  $\xi_{g,R}$ are positively correlated. 

Without fixing either the curvature of the universe  (flat prior $0<|\Omega_k|<1$) or the quality 
of the DE component (flat prior $-3/2<w<-1/2$),  but taking (strong) Gaussian priors of  $\Omega_b h^2$, $H_0$, and $n_s$ 
from big bang nucleosynthesis (BBN, \cite{bbn}), Hubble Space Telescope (HST, \cite{hst}),
and WMAP7 \citep{lars}, respectively, we constrain the local abundance of matter $\Omega_m$ 
with   a  precision of  $8\%$ ($\Omega_m=0.283\pm0.023$, see Fig. \ref{fig:data}.
For the sake of completeness,  by choosing a smoothing scale of $18/22h^{-1}$Mpc, 
we obtain $\Omega_m=0.270^{+0.037}_{-0.027}/0.255 \pm 0.038$.)
This figure improves the precision of the estimate  ($\Omega_m=0.259\pm 0039$) by a factor of two obtained by analysing, with the same priors,  
the BAO of the SDSS DR5 sample (which contains 
25\% fewer galaxies than DR7)\citep{eis}. It also improves by $\sim 20\%$ 
the precision of the constraint $\Omega_m=0.24^{+0.025}_{-0.024}$ that  \cite{bao} obtained by combining 
BAO results from the DR7 sample with the  full  likelihood of the WMAP5 data \citep{cmb5} and a strong HST prior \citep{riess09}.
The value  $\Omega_m=0.294\pm0.017$ \citep{anderson} obtained by combining the
BAO results from the DR9 sample  (containing 5 times more LRGs than the DR7 catalogue) with the full likelihood of CMB data \citep[WMAP7, ][]{lars} is nearly $40\%$ more precise than our estimate.

The best-fitting value  of $\Omega_m$  still remains  within the
quoted $68\%$ contour level (cl), even when we 
relax some of the strong priors. With a (weak) flat prior on $n_s$ (in the interval 
$[0.9,1.1]$), we obtain $\Omega_m=0.271^{+0.030}_{-0.031}$, while if
we go on to 
weaken the prior  on  $\Omega_bh^2$  (flat in the interval $[0,0.03]$), we obtain  
$\Omega_m=0.255\pm 0.040$. Although the best-fitting value of $\Omega_m$ is weakly 
sensitive to changes in $\Omega_b h^2$ and $n_s$, Fig. \ref{fig:degeneracy} shows 
that $\Omega_m$  degenerates with $H_0$ when the HST prior is removed.
If we also relax the strong prior on $H_0$ (by assuming a 
flat prior in the interval $[40,100]$km/s/Mpc), we find $\Omega_m=0.22^{+0.14}_{-0.04}$. 
The stability of the best-fitting central value $\Omega_m$ is of even more 
interest if contrasted to CMB results showing  that it is the combination  
$\Omega_m h^2$  that is insensitive to the prior on the curvature of the universe. 
On the contrary, if on top of the chosen priors we also  impose as constraints  that 
the universe is flat and that dark energy is effectively described by a cosmological 
constant  (i.e. we fix $w=-1$),  we obtain $\Omega_{m}=0.275\pm0.020$, an estimate  that improves
the WMAP7 bound by  $50 \%$ ($0.267\pm0.029$, \citet{lars}). For comparison, the recent combination of the Planck temperature power spectrum 
with WMAP polarization gives $\Omega_{m,0}=0.315\pm0.017$.

By switching from precision to accuracy, it is important to point out that 
the mass-clustering ratio (cf. Eq. \ref{etar})  computed using  the best parameters ${\bf p}$ inferred
on a scale $R=14h^{-1}$Mpc correctly predicts  
the galaxy clustering ratio observed on various different scales $R$ (see lower panel of Fig. \ref{fig:eta}).
This highlights the overall unbiasedness of our cosmological inference.

The Fisher matrix formalism cannot  be reliably applied to reproduce  constraints on cosmological parameters that are strongly degenerate. 
A simple way to make sense and reproduce our results consists of working  out  an effective measure of the observable $\eta_{g,R}(r, {\bf p})$.  
The relevant cosmological information contained in the SDSS DR7 data  can be effectively retrieved by approximating the galaxy clustering ratio $\eta_{g,R}$ 
via the fitting formula 

\begin{equation}
\eta_{g,R}(3R,{\bf p})\!=\!0.0712\!-\!0.0788 x_p\!+\!0.0981 x^2_p\!-\!0.0538x^3_p
\label{appeta}
\end{equation}
\noindent where $R=14h^{-1}$Mpc, $x_p=1+{\mathrm f}_p/8.48$, and where
$$
\begin{array}{rl}
{\mathrm f}_p\!=\!&\!3w\Omega_X\!-\!\Omega_k\!+0.042\left\{w\Omega_X[1\!-\!\frac{9}{2}(1+w)w]\!+\!\frac{\Omega_k^2-5\Omega_k}{2}\right\}\\
                   & +0.29\left\{w\Omega_X(14+3w)-3w\Omega_X\Omega_k+\frac{\Omega_k^2}{2}-\frac{11}{3}\Omega_k\right\}.
\end{array}
$$
\noindent

This  phenomenological formula  allows a fast computation of  the  value of  the observable $\eta_{g,R}$ in any given 
cosmological model characterized by  $0.1 < \Omega_m< 1$, $0<\Omega_X<1$,  and $-1.5 < w < -0.5$.
The left-hand panel of Fig. \ref{fig:appk} is obtained by  setting $w=-1.2$ and by taking isocontours of the resulting 
two-dimensional surface  in the plane $[\Omega_m, \Omega_X]$.  By  inserting this formula in the  likelihood expression (\ref{logpost}), and by further   
assuming $\sigma_{\eta}=0.004$ independently of the cosmological model, one  can 
reproduce the  cosmological constraints shown in Fig. \ref{fig:data}.

\section{Conclusions}

The paper presents and illustrates the use of  a new cosmic standard
of clustering, the galaxy clustering ratio $\eta_{g,R}(r)$, which is
defined  as the ratio of the correlation function to the variance of the  galaxy overdensity field smoothed on a scale $R$.
The key result of  the analysis, the one that  a-posteriori justifies by itself  the introduction of  this  statistic,  is the demonstration  that 
there is no need to model  the  complex physics  underpinning galaxy formation and evolution processes if we are to
deduce the  amplitude of the corresponding mass statistics ($\eta_R(r)$) from the observed value of the galaxy clustering ratio ($\eta_{g,R}(r)$) . 
Indeed,  it is straightforward to show that, on large linear scales,  $\eta_{g,R}(r)=\eta_R(r)$. In other words,   
what you see (baryon clustering) is what you get (dark matter clustering).

Interestingly,  the magnitude of  the  mass clustering ratio $\eta_R(r)$ on the given fixed scales R and r is a number that  is neatly related to the relative amplitude of the 
the  real-space mass power spectrum  in two distinct bandwidths. 
This implies that, 
instead of being characteristic  of  a specific  galaxy sample, the galaxy clustering ratio estimated for the pair of linear scales R and r  is a universal quantity that only depends 
on the  initial distribution of matter density fluctuations and on the nature of dark matter. 

Many other cosmological observables probe information encoded in second-order mass statistics. 
The distinctiveness of  the galaxy clustering ratio  resides primarily  in its  simplicity.  On large linear scales,  its amplitude is  independent of  galaxy biasing,  
peculiar velocity modelling,  linear growth rate of structures, and normalization of the matter  power spectrum. 
The advantage over standard clustering analyses is both practical and conceptual.  The clustering ratio provides a way to
extract information about the mass power spectrum without the need
 to estimate  the galaxy power spectrum
or the correlation function of galaxies.   Moreover, that  $\eta_{g,R}(r)$ is estimated via  cell-counting  techniques 
on a given fixed pair of scales $(R,r)$ makes  the estimation process and the  likelihood analysis  computationally  fast. 
There is no need, for example, to compute the  sample covariance matrix or  to speed up calculations by framing the clustering analysis
in a given fiducial cosmological model. Additionally, since results do not follow from the reconstruction of the full shape of the 
galaxy correlation function $\xi(s)$,  the cosmological interpretation is in principle less affected  by observational biases that act differentially 
as a function of scale $s$.  The $\eta$ formalism is  also conceptually transparent. Only a minimum number of physical hypotheses, and no astrophysical 
nuisance parameters  at all, condition the cosmological inference:  the clustering ratio   can  be extracted from data and  compared to theoretical 
predictions  or numerical simulations  in a fairly straightforward way.

The method's robustness  is thoroughly tested via  an end-to-end  analysis  of numerical simulations of  the large scale structure of the 
universe. We  have shown that the galaxy density field can be smoothed and correlated on opportunely chosen scales 
 $R$ and $r$ so that  the {\it galaxy}  clustering ratio effectively   measured in simulations  and  the {\it mass} clustering ratio  predicted by  theory  differs,  at most, by  $0.1\%$ on large linear scales $R$. 
This  remarkable agreement explains both  the precision and the accuracy of the method in retrieving, by means of  a blind analysis,  the values of the matter density parameter $\Omega_m$  
of various numerical simulations of the large scale structure of the universe.

We have also demonstrated the method with real data. Using the LRG sample extracted from  the spectroscopic SDSS
 DR7 galaxy catalogue,  no CMB  information, weak (flat) priors  on the value of the 
curvature  of the   universe ($\Omega_k$)  and the  constant value of
the dark energy equation of state ($w$), we find
$\Omega_m=0.271^{+0.030}_{-0.031}$. Since one of the main goals of the
paper was  to illustrate  the intrinsic strengths and limitations of
the $\eta-$test, we did not present cosmological results  from a full
joint analysis with CMB  data. This analysis, which will allow priors
to be lifted and a broader range of cosmological parameters to be constrained including $h$ and $n_s$, is left  for a future work. 
    
Of strong interest is the flexibility of the method, which can still  be improved along several directions. Owing to its scale-free nature, 
the precision of the technique could be further improved  by adopting a non-linear power spectrum in eqs. (\ref{ff}) and (\ref{gg}), 
and by smoothing the over-density field  on an even smaller scale $R$
than the one adopted
throughout this analysis.  There are  a few caveats to this approach, though.  One must verify that mass fluctuations $\delta_R$ in
real  and redshift spaces  are still approximately proportional, as predicted, for example, by
the  linear Kaiser model.  More importantly, one must  properly describe the non-linear redshift space
distortions induced by virial motions of  galaxies, the so called
Finger-of-God effect. As pointed out by \citet{neyrinck} ,  who developed a similar statistic in Fourier space, this
is a highly non-linear phenomenon  that is expected to be significant  on small smoothing scales $R$. 
A strategy for overcoming these difficulties and implementing the $\eta$ test even on such extreme regimes is  
explained and applied to the VIPERS \citep{guz13}  high-redshift spectroscopic data by \citet{belma}.
Alternatively, the method could also be  applied to large photometric redshift  surveys. 
In this case, the predictions of  eq. (\ref{etar})  must be statistically corrected to account for the line-of-sight distortions 
introduced when estimating low-resolution distances from photometry. Finally, this probe enriches the arsenal of methods with which the next generation 
of redshift surveys such as EUCLID and BigBOSS will hunt  for new physics by  challenging all sectors of the cosmological model.

\begin{acknowledgements}
We acknowledge useful discussions with F. Bernardeau,  E. Branchini, E. Gazta\~naga, B. Granett, L. Guzzo, Y. Mellier, L. Moscardini, 
A. Nusser, J. A. Peacock, W. Percival,  H. Steigerwald, I. Szapudi,
and P. Taxil. CM is grateful for support from  specific project funding of the {\it Institut Universitaire de France} and of the Labex OCEVU. 

\end{acknowledgements}

\end{document}

%% file: macros.tex
\def\lsim{\raise0.3ex\hbox{$<$}\kern-0.75em{\lower0.65ex\hbox{$\sim$}}} 
\def\gsim{\raise0.3ex\hbox{$>$}\kern-0.75em{\lower0.65ex\hbox{$\sim$}}} 
\def\lesssim{\mathrel{\hbox{\rlap{\hbox{\lower4pt\hbox{$\sim$}}}\hbox{$<$}}}}
\def\gtrsim{\mathrel{\hbox{\rlap{\hbox{\lower4pt\hbox{$\sim$}}}\hbox{$>$}}}}
\def\sc1{\raise2.1ex\hbox{\tiny $r\!\!=\!\!4$}\kern-0.95em{\hbox{$=$}}} 
\def\ltsima{$\; \buildrel < \over \sim \;$}
\def\simlt{\lower.5ex\hbox{\ltsima}}
\def\gtsima{$\; \buildrel > \over \sim \;$}
\def\simgt{\lower.5ex\hbox{\gtsima}}          

\newcommand{\unit}[1]{\ifmmode \:\mbox{\rm #1}\else \mbox{#1}\fi}

\def\sc{{\rm Science\ }}

\def\apjl{ApJL}

\def\aap{A\&A}

\def\apjs{ApJ. S}

\def\gsim{~\rlap{$>$}{\lower 1.0ex\hbox{$\sim$}}}

\def\simpropto{\lower.2ex\hbox{$\; \buildrel \propto \over \sim \;$}}
\def\ltsim{\lower.5ex\hbox{$\; \buildrel < \over \sim \;$}}
\def\gtsim{\lower.5ex\hbox{$\; \buildrel > \over \sim \;$}}
\def\ltsim{\lower.5ex\hbox{$\; \buildrel < \over \sim \;$}}
\def\gtsim{\lower.5ex\hbox{$\; \buildrel > \over \sim \;$}}

\def\ln{{\rm ln}}

\def\pmb#1{\setbox0=\hbox{#1}%
\kern-.025em\copy0\kern-\wd0
\kern.05em\copy0\kern-\wd0
\kern-.025em\raise.0433em\box0}

\def\vr{\pmb{$r$}}

\def\simlt{\lower.5ex\hbox{$\; \buildrel < \over \sim \;$}}
\def\simgt{\lower.5ex\hbox{$\; \buildrel > \over \sim \;$}}

\newcommand{\beq}{\begin{equation}}
\newcommand{\eeq}{\end{equation}}
\def\beqa{\begin{eqnarray}}
\def\eeqa{\end{eqnarray}}
\def\fixit#1{}

%% file: definitions.tex
\def\abc2{{\left(\frac{ab}{c^2}\right)}}

\def\tpi3{(2\pi)^3}

\def\be{\begin{equation}}
\def\ee{\end{equation}}

\def\simless{\mathbin{\lower 3pt\hbox
  {$\rlap{\raise 5pt\hbox{$\char'074$}}\mathchar"7218$}}}   % < or of order
\def\simgreat{\mathbin{\lower 3pt\hbox
   {$\rlap{\raise 5pt\hbox{$\char'076$}}\mathchar"7218$}}}  % > or of order

\def\ba{\begin{eqnarray}}
\def\ea{\end{eqnarray}}

\def\bess0{{\rm J_0}}
\def\sbess0{{\rm j_0}}

\def\1H{1{\rm H}}
\def\2H{2{\rm H}}
\def\3H{3{\rm H}}

\def\D1{\rm 1D}

\def\vr{{\hbox{\BF r}}}

\def\vr{{\bf r}}

\def\bce{\begin{center}}
\def\ece{\end{center}}
\def\ben{\begin{enumerate}}
\def\een{\end{enumerate}}

\def\brr{\begin{array}}
\def\err{\end{array}}

\def\beq#1{\begin{equation}#1\end{equation}}
\def\be{\begin{equation}}
\def\ee{\end{equation}}
\def\bea{\begin{eqnarray}}
\def\eea{\end{eqnarray}}

\def\ba{\begin{eqnarray}}
\def\ea{\end{eqnarray}}

\def\fun#1#2{\lower3.6pt\vbox{\baselineskip0pt\lineskip.9pt
        \ialign{$\mathsurround=0pt#1\hfill##\hfil$\crcr#2\crcr\sim\crcr}}}

\def\spose#1{\hbox to 0pt{#1\hss}}

%% file: ms.bbl
\begin{thebibliography}{99}

\bibitem[\protect\citeauthoryear{Abazajian et al.}{2009}]{sdss} Abazajian, K. N., Adelman-McCarthy, J. K., Agueros, M. A., et al. 2009,  ApJS, 182, 543 

\bibitem[\protect\citeauthoryear{Ade et al.}{2013}]{papPlanck} Ade, et al. 2013, A\&A submitted

\bibitem[\protect\citeauthoryear{Alcock \& Paczynski}{1979}]{ap} Alcock, C. \&  Paczynski, B. 1979, Nature, 281, 358  

\bibitem[\protect\citeauthoryear{Amendola et al.}{2012}]{amen12} Amendola, L., Appleby, S., Bacon, D., et al. 2012, arXiv:1206.1225

\bibitem[\protect\citeauthoryear{Anderson et al.}{2012}]{anderson} Anderson, L., Aubourg, E. Bailey, S.  et al. 2012, MNRAS, 427, 3435
 
\bibitem[\protect\citeauthoryear{Ballinger, Peacock \&  Heavens}{1996}]{bph}  Ballinger, W. E., Peacock, J. A. \&  Heavens, A. F. 1996, MNRAS, 282,  877

\bibitem[\protect\citeauthoryear{Bel \& Marinoni}{2012}]{bm}  Bel, J. \& Marinoni, C. 2012, MNRAS, 424, 971

\bibitem[\protect\citeauthoryear{Bel et al.}{2013}]{belma} Bel, J., Marinoni,C., Granett, B., et al. (the VIPERS collaboration) 2013, A\&A accepted, arXiv:1310.3380B

\bibitem[\protect\citeauthoryear{Bernardeau}{1996}]{berna}  Bernardeau, F. 1996, A\&A, 312, 11

\bibitem[\protect\citeauthoryear{Bernardeau et al.}{2002}]{berna02} Bernardeau, F., Colombi, S., Gazta\~naga, E. \& Scoccimarro, R. 2002, Phys. Rept., 367, 1 

\bibitem[\protect\citeauthoryear{Blake et al.}{2012}]{mod3} Blake, C., Brough, S., Colless, M., et al. 2012, arXiv:1204.3674  

\bibitem[\protect\citeauthoryear{Clifton et al.}{2012}]{clif} Clifton, T., Ferreira, P. G., Padilla, A. \& Skordis, C. 2012, Physics Reports, 513, 1

\bibitem[\protect\citeauthoryear{Copeland, Sami \&  Tsujikawa}{2006}]{cope}   Copeland,  E. J.,  Sami, M. \&  Tsujikawa, S. 2006,  International Journal of Modern Physics D, 15, 1753

\bibitem[\protect\citeauthoryear{Dunkley et al.}{2009}]{cmb5} Dunkley, J., Spergel, D.N., Komatsu, E., et al. 2009, ApJS, 180, 306

\bibitem[\protect\citeauthoryear{Eisenstein \&  Hu}{1998}]{EH}  Eisenstein, D. \&  Hu, W. 1998, ApJ, 496, 605

\bibitem[\protect\citeauthoryear{Eisenstein et al.}{2005}]{eis} Eisenstein, D., Zehavi, I., Hogg, D.W., et al. 2005, ApJ, 633, 560

\bibitem[\protect\citeauthoryear{Feng}{2010}]{feng} Feng, J. 2010. ARA\&A  48, 495

\bibitem[\protect\citeauthoryear{Frenk \& White}{2012}]{fw}  Frenk, C. S. \&  White, S. D. M. 2012, Ann. Phys.,  524, 507

\bibitem[\protect\citeauthoryear{Frieman, Turner  \&  Huterer}{2008}]{frie} Frieman, J. A., Turner, M. S.  \&  Huterer, D.  2008, ARA\&A, 46, 385

\bibitem[\protect\citeauthoryear{Fry \& Gazta\~{n}aga}{1993}]{fg}  Fry, J. N.  \& Gazta\~{n}aga, E. 1993, ApJ, 413, 447 

\bibitem[\protect\citeauthoryear{Gaztanaga \& Cabre}{2009}]{bao2}  Gaztanaga, E. \&  Cabre, A. 2009, MNRAS, 399, 1663

\bibitem[\protect\citeauthoryear{Guzzo et al.}{2008}]{guz} Guzzo, L., Pierleoni, M., Meneux, B., et al. 2008, Nature, 451,541

\bibitem[\protect\citeauthoryear{Guzzo et al.}{2013}]{guz13} Guzzo, L., Scodeggio, M. Garilli, B. et al. 2013, arXiv:1303:2623 

\bibitem[\protect\citeauthoryear{Jenkins et al.}{2001}]{HVP} Jenkins, A.,  Frenk, C. S., White, S. D. M., et al. 2001, MNRAS, 321, 372

\bibitem[\protect\citeauthoryear{Kaiser}{1987}]{kaiser} Kaiser, N. 1987,  MNRAS, 227, 1

\bibitem[\protect\citeauthoryear{Komatsu  et al.}{2011}]{mod1}  Komatsu, E., Smith, K.M., Dunkley, J., et al. 2011, ApJS, 192, 18 

\bibitem[\protect\citeauthoryear{Larson et al.}{2011}]{lars} Larson, D.,  Dunkley, J., Hinshaw, G., et al. 2011, ApJS,192, 16

\bibitem[\protect\citeauthoryear{Laureijs  et al.}{2011}]{euc} Laureijs, R., Amiaux, J., Arduini, S., et al. 2011, arXiv:1110.3193

\bibitem[\protect\citeauthoryear{Layser}{1956}]{Layser} Layser, D. 1956, AJ, 61, 1243

\bibitem[\protect\citeauthoryear{Le F\`evre et al.}{2005}]{lefevre} Le F\`evre, O., Vettolani, G., Garilli, B., et al. 2005, A\&A, 439, 845 

\bibitem[\protect\citeauthoryear{ Marinoni  et al.}{2005}]{marinoni}  Marinoni, C., Le F\`evre, O., Meneux, B., et al. 2005, A\&A, 442, 801 

\bibitem[\protect\citeauthoryear{Marinoni et al.}{2008a}]{mardiscs} Marinoni, C., Saintonge, A., Giovanelli, R., et al. 2008, A\&A, 478, 43 

\bibitem[\protect\citeauthoryear{Marinoni  et al.}{2008b}]{marinoni08}  Marinoni, C. Guzzo, L.,  Cappi, A. et al. 2008, A\&A, 487, 7
 
\bibitem[\protect\citeauthoryear{Marinoni \& Buzzi}{2010}]{marpairs} Marinoni, C. \& Buzzi, A.  2010, Nature, 468, 539 

\bibitem[\protect\citeauthoryear{Marinoni, Bel  \& Buzzi}{2012}]{mariso}   Marinoni, C. Bel, J. \& Buzzi, A. 2012, JCAP, 10, 036
 
\bibitem[\protect\citeauthoryear{Martin}{2012}]{jer2}  Martin, J.  2012, arXiv:1205.3365

\bibitem[\protect\citeauthoryear{McBride et al.}{2009}]{lasdamas} McBride, C.,  Berlind, A., Scoccimarro, R., et al. 2009, AAS, 21342506

\bibitem[\protect\citeauthoryear{Neyrinck}{2011}]{neyrinck} Neyrinck,  M. 2011, ApJ, 742, 91

\bibitem[\protect\citeauthoryear{Peebles \& Ratra}{2003}]{peeb} Peebles, P. J. E.  \& Ratra, B.  2003, Rev. Mod. Phys., 75, 559

\bibitem[\protect\citeauthoryear{Percival et al.}{2010}]{bao}  Percival, W. J., Reid, B.A., Eisenstein, D.J., et al. 2010, MNRAS, 401, 2148

\bibitem[\protect\citeauthoryear{Pettini et al.}{2008}]{bbn}  Pettini, M., Zych, B., Murphy, M.T., Lewis, A. \& Steidel, C.C. 2008, MNRAS, 391, 1499

\bibitem[\protect\citeauthoryear{Reid et al.}{2012}]{mod4} Reid, B. A., Samushia, L., White, M., et al. 2012,  MNRAS, 426, 2719

\bibitem[\protect\citeauthoryear{Riess et al.}{2009}]{riess09} Riess, A. G., Macri, L., Casertano, S., et al. 2009, 699, 539

\bibitem[\protect\citeauthoryear{Riess et al.}{2011}]{hst}  Riess, A., Macri, L., Casertano, S., et al. 2011, ApJ, 730, 119

\bibitem[\protect\citeauthoryear{Scherrer \& Weinberg}{1998}]{scherrer}  Scherrer, R. J. \& Weinberg, D. H. 1998, ApJ, 504, 607

\bibitem[\protect\citeauthoryear{Sanchez, et al.}{2012}]{san}  Sanchez, A. G., Sc\'occola, C. G., Ross, A. J., et al. 2012, MNRAS, 425, 415

\bibitem[\protect\citeauthoryear{Schlegel et al.}{2011}]{boss}  Schlegel, D., Abdalla, F., Abraham, T., et al. 2011,  arXiv1106.1706 

\bibitem[\protect\citeauthoryear{Smith et al.}{2003}]{smith} Smith, R. E., Peacock, J. A., Jenkins, A., et al. 2003, MNRAS, 341, 1311

\bibitem[\protect\citeauthoryear{Szapudi, Szalay \& Bosch\'an}{1992}]{szasza1} Szapudi, I., Szalay, A. \& Bosch\'an, P. 1992, ApJ, 390, 350

\bibitem[\protect\citeauthoryear{Szapudi \& Szalay}{1997}]{szasza2} Szapudi, I. \&  Szalay, A. 1997, ApJ, 481, 1

\bibitem[\protect\citeauthoryear{Szapudi \& Szalay}{1998}]{szasza3} Szapudi, I. \&  Szalay, A. 1998, ApJ, 494, 41

\bibitem[\protect\citeauthoryear{Weinberg}{1989}]{jer1}  Weinberg, S. 1989, Rev. Mod. Phys., 61, 1

\bibitem[\protect\citeauthoryear{Weinberg}{2008}]{cosmo}  Weinberg, S. 2008, Cosmology,  Oxford University Press

\end{thebibliography}
